\documentclass[amsmath,amssymb,aps,floatfix,superscriptaddress,twocolumn,nofootinbib]{revtex4-2}

\usepackage{graphicx}
\usepackage[caption=false]{subfig}
\usepackage{dcolumn}
\usepackage{bm}
\usepackage{braket}
\usepackage{appendix}
\usepackage{todonotes}
\usepackage{soul}
\usepackage[colorlinks = true, linkcolor = blue, citecolor = blue, urlcolor = blue]{hyperref}
\usepackage{academicons}
\usepackage{minted}

\newminted{python}{%
}

\newcommand{\orcid}[1]{\href{https://orcid.org/#1}{\textcolor[HTML]{A6CE39}{\aiOrcid}}}

\begin{document}

\preprint{APS/123-QED}

\title{Multi-stage Quantum Amplifier Readout Chain}

\author{L.~Howe}
\email{lhowe@caltech.edu}
\affiliation{California Institute of Technology, Pasadena, California 91125, USA}
\affiliation{National Institute of Standards and Technology, Boulder, 80305, Colorado, USA}
\affiliation{Department of Physics, University of Colorado, Boulder, 80309, Colorado, USA}

\author{A.~Giachero}
\email{andrea.giachero@unimib.it}
\affiliation{National Institute of Standards and Technology, Boulder, 80305, Colorado, USA}
\affiliation{Department of Physics, University of Colorado, Boulder, 80309, Colorado, USA}
\affiliation{Department of Physics, University of Milano-Bicocca, Milan, I-20126, Italy}

\author{M.~Vissers} 
\affiliation{National Institute of Standards and Technology, Boulder, 80305, Colorado, USA}

\author{C.~Shiu} 
\affiliation{National Institute of Standards and Technology, Boulder, 80305, Colorado, USA}

\author{S.~Duff}
\affiliation{National Institute of Standards and Technology, Boulder, 80305, Colorado, USA}

\author{J.~Austermann} 
\affiliation{National Institute of Standards and Technology, Boulder, 80305, Colorado, USA}

\author{J.~Hubmayr}
\affiliation{National Institute of Standards and Technology, Boulder, 80305, Colorado, USA}

\author{J.~Ullom}
\affiliation{National Institute of Standards and Technology, Boulder, 80305, Colorado, USA}
\affiliation{Department of Physics, University of Colorado, Boulder, 80309, Colorado, USA}

\date{\today}

\begin{abstract}
Multi-stage cryogenic readout chains with a wide bandwidth and added noise within a few quanta of the quantum limit are frequently constructed using traveling-wave parametric amplifiers (TWPAs) as the first stage, and a semiconductor amplifier as the second stage. Unfortunately, for highly-scaled superconducting detector arrays, or quantum information systems, and space-based observatories, the power dissipation of the semiconductor amplifier becomes problematic from the perspective of available cryogenic cooling power near 3~K. Here we demonstrate a readout chain based on a two-stage kinetic inductance TWPA (KTWPA). This quantum-amplifier-based-readout-chain (QARC) provides sufficient gain that a cryogenic semiconductor follow-on amplifier can be eliminated without degradation of the system noise. Due to the ultra-low dissipation of KTWPAs and their higher output-power-to-total-power ratio, the QARC dissipates approximately three orders of magnitude less power than readout chains containing semiconductor amplifiers while adding noise of less than 2~quanta over a 1~GHz bandwidth. In addition, by leveraging the high power handling of kinetic inductance technology, the QARC maintains an input compression point of -93~dBm, which exceeds that of many contemporary Josephson-junction-based parametric amplifiers.
\end{abstract}

\maketitle



\section{Introduction}
Efficient, low-dissipation, and low-noise amplifiers are essential for the readout of superconducting detectors and quantum devices. Ideally one is able to construct a multi-stage amplifier readout chain capable of nearly-quantum-limited (nQL) added noise, with a wide bandwidth and large dynamic range. Contemporary nQL readout chains universally employ a superconducting parametric quantum amplifier \cite{yurke1989observation, krantz2019quantum, malnou2024low} as the first stage at sub-Kelvin temperatures, and a semiconductor amplifier as the second stage at \mbox{3~K to 4~K}. Using a traveling wave parametric amplifier (TWPA) \cite{macklin2015near, malnou2021three}, rather than a resonant quantum amplifier \cite{castellanos2007widely}, at the first stage has enabled demonstrations of multi-gigahertz bandwidths with added noise between \mbox{1~quanta and 3~quanta} \cite{howe2025kinetic, faramarzi2024kinetic, malnou2024low}. However, the cryogenic power dissipation of these readout chains is dominated by the semiconductor amplifier and is significant enough to hamper realization of systems requiring ultra-low dissipation or high levels of scaling, such as complex quantum information experiments \cite{mohseni2024build}, or space-based observatories leveraging large superconducting detector arrays \cite{van2025power, bradley2021advancements, dipirro2019lynx, bandler2019lynx}.

TWPAs require a reactive nonlinearity and most superconducting incarnations utilize the Josephson junctions and their associated nonlinear inductance \cite{macklin2015near, ranadive2022kerr, planat2020photonic, qiu2023broadband, simbierowicz2021characterizing, gaydamachenko2025rf}. While Josephson TWPAs (JTWPAs) are the predominant TWPA technology employed in superconducting quantum information systems they suffer from magnetic field sensitivity \cite{janssen2024magnetic}and limited dynamic range. Kinetic inductance TWPAs (KTWPAs) instead rely on the nonlinear superconducting kinetic inductance and are a burgeoning technology with the potential to address these limitations \cite{howe2025kinetic, malnou2021three, faramarzi2024kinetic, boutan2026operation, janssen2026temperature} while easing fabrication demand \cite{kissling2024vulnerability}. Recently, KTWPAs have demonstrated high fidelity readout of long coherence time qubits -- with no quantifiable backaction or heating \cite{castellanos2025measureable} -- and have been demonstrated at 4~K as a potential low-power replacement for HEMT amplifiers \cite{malnou2022performance} ubiquitous in cryogenic microwave readout.

\begin{figure*}[t!]
    \centering
    \includegraphics[width = 0.98\textwidth]{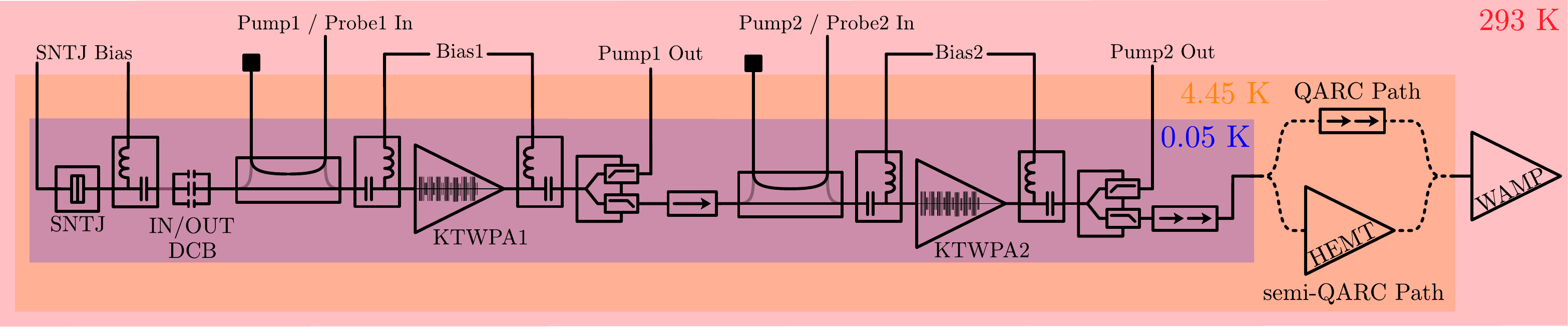}
    \caption{Simplified experimental schematics for the QARC and semi-QARC which we characterize in separate cooldowns in a dilution refrigerator. Each KTWPA is individually biased and pumped for maximum tunability and both individual and in-tandem performance optimization. Diplexers are used at the output of each KTWPA to pick off the residual pump and avoid compression of the upstream amplifier. 4~GHz to 12~GHz isolators are used between each stage to enforce directionality in the readout chain. A single-stage isolator is located at the KTWPA1 output, a double stage at the KTWPA2 output and, in the case of the QARC setup, the HEMT is replaced with an additional double-stage isolator to replicate the reverse-isolation provided by the HEMT. The KTWPA pumps and vector network analyzer (VNA) probe tones are combined at room temperature and injected at each KTWPA input at 0.05~K using directional couplers. Calibration of the loss of all components in the dual-KTWPA chain is performed using a separate bypass cable (not shown). A shot-noise tunnel junction (SNTJ) is located at the QARC/semi-QARC input (downstream of KTWPA1) and provides a calibrated white noise to permit characterization of the QARC/semi-QARC system noise at the KTWPA1 input reference plane. An inner-outer dc block (DCB) is used to enable differential biasing of the SNTJ by breaking the galvanic connection between the SNTJ base electrode and the cryostat body. DC biasing of the KTWPAs is also done differentially using isolated battery voltage sources at 300~K.}
    \label{fig:double_kit_setup}
\end{figure*}

In the aforementioned scaled or cooling-power-starved situations it is beneficial to consider cryogenic amplification chains consisting only of quantum (parametric) amplifiers -- such as the system studied in our work depicted in Fig.~\ref{fig:double_kit_setup}. These quantum-amplifier-based readout chains (QARCs) are achieved via removal or replacement of the semiconductor amplifiers typically located near 3~K \cite{malnou2022performance, bandler2019lynx, dipirro2019lynx, bradley2021advancements, van2025power}. See Fig.~\ref{fig:double_kit_setup} for a simplified experimental schematic of our implementation. In an ideal case such systems possess sufficient parametric gain so as not to rely on transistor-based gain to realize an amplifier chain with nQL performance. The primary advantages offered by quantum amplifiers such as KTWPAs are (1) access to nQL amplification, and (2) substantially lower power dissipation compared to semiconductor amplifiers. Three-wave-mixing KTWPA dissipation at the base cryogenic stage can be estimated by assuming a reasonable off-chip bias network resistance from the cold finger to the device (0.5~$\Omega$), pump frequency (12~GHz) and by considering a typical frequency-dependent loss of 1~dB/GHz for the KTWPA itself \cite{macklin2015near, ranadive2022kerr, ranadive2024traveling, planat2020photonic, gaydamachenko2025rf, qiu2023broadband, simbierowicz2021characterizing}. Under a dc bias of 0.5~mA and with a pump power of -40~dBm at the package input \cite{howe2025kinetic} our KTWPAs dissipate approximately 120~nW of dc power and 90~nW of rf power at their mounting stage. Properly thermalizing the pump injection line for a quantum information system requires approximately 40~dB -- typically with 20~dB at the 3~K stage and 20~dB either at the mixing chamber base (0.01--0.02~mK) or precooling stage (0.05--0.1~mK). See Fig.~\ref{fig:full_schematic} for our complete implementation. Requiring -40~dBm of pump power at the KTWPA input places pump dissipation at the 3~K stage at $10~\mu$W. 

A state-of-the-art HEMT dissipates three orders of magnitude more power: typically \mbox{10~mW--15~mW} \cite{lnf, van2025power}. However, this power dissipation is at the \mbox{3~K--4~K} where cooling powers for \textit{terrestrial} systems are on the order of Watts. Nonetheless, the scaling demands for quantum error correction quickly result in HEMT quantities whose dissipation exceeds the capacity of a single mechanical cryocooler \cite{zeng2023low, saadatmand2026superconducting, malnou2022performance}. Depending on qubit multiplexing factor this equality is reached for $\mathcal{O}(10^3-10^4)$ qubits. The error correction threshold requires multiple orders of magnitude larger quantum processors \cite{fowler2021surface, stephens2014fault} -- and simply adding additional cryocoolers only scales linearly while displacing valuable space required for signal input/output cabling. Importantly, our KTWPA dissipation exceeds dilution refrigerator cooling budgets of $\sim 10~\mu$W at 0.02~K at approximately the same scaling. This indicates that current KTWPAs, while still substantially lower power than HEMTs, are particularly well-suited for HEMT-elimination via operation at a non-mixing-chamber stage in scaled quantum systems.

Space-based missions \cite{bandler2019lynx, dipirro2019lynx, durand2008mid, moore2017mid, glenn2025prima, odagiri2022cryogenic} occupy a different design space; with mechanical cryocooler base stage temperatures in the \mbox{4~K--6~K} range supporting only \mbox{20~mW--70~mW} of static dissipation. Thus even a single HEMT dominates space-based cooling budgets -- with certain experiments being forced to increase readout system noise, by reducing HEMT bias and thus gain, in order to operate their detector arrays \cite{bandler2019lynx}. These aspects indicate that pure QARCs can enable significantly lower dissipation than semiconductor-assisted QARCs (semi-QARCs).

For a QARC to achieve nQL operation at least 40~dB of gain is required. This  amount of gain takes the QL of 0.5~quanta (0.24~K at 5~GHz) and amplifies it above 300~K. Since no single KTWPA has demonstrated these levels of gain \cite{howe2025kinetic, malnou2021three, faramarzi2024kinetic} we instead seek to achieve the requisite gain using a double-KTWPA setup -- each with approximately 20~dB of gain. Similar multi-stage quantum amplifier readout chains have been demonstrated previously \cite{lilja2026phase, lilja2026performance, vaartjes2024strong, lecocq2021efficient, mallet2011quantum} but with either insufficient gain, or not necessarily with the aim of semiconductor amplifier elimination. To ease integration in our experimental setup we choose to locate both KTWPA stages at the mixing chamber base stage of a dilution refrigerator -- with the knowledge that operation of the second-stage KTWPA at higher temperature stages \cite{malnou2022performance} is not only possible but also enhances scalability.

Note that, in our multi-stage KTWPA case where nominally identical amplifiers are used, this has the effect of moving the amplifier chain input compression point down by the gain of the first KTWPA. In future double-KTWPA systems this reduction in the chain input compression can be avoided if the second amplifier's dynamic range is higher than that of the first by a factor of the first stage's gain. This is easily accomplished by increasing the film cross-sectional area by a factor of two to four; albeit with a commensurate increase in dissipation. Prior KTWPAs \cite{malnou2021three} had twice the area as the KTWPAs studied herein \cite{howe2025kinetic, howe2025compact, giachero2024kinetic} and featured correspondingly higher compression points.

A salient efficiency metric for comparing amplifiers is the ratio of the output compression power to the amplifier's total power budget:
\begin{equation}
    \beta = \frac{OP_1}{P_\text{tot}} = \frac{OP_1}{P_\text{diss} + OP_1}.
\end{equation}
Here, $OP_1$ is the output 1~dB compression point, and $P_\text{diss}$ is the total dissipated power on the stage at which the amplifier resides. This ratio quantifies how efficiently any given amplifier converts energy from the applied bias reservoir to output power in the signal band -- which in the end is what an experimenter is interested in. For an ideal, lossless amplifier $\beta$ approaches unity and all bias provided to the amplifier (dc bias and pump) is perfectly converted to output signal power with no dissipation. State-of-the-art HEMT amplifiers \cite{lnf} feature output compression points of approximately -5~dBm and dissipate 15~mW of dc power, so $\beta_\text{HEMT} \sim 0.02$. Recent experimental low-power HEMTs \cite{li2024investigation, cha2023optimization} have been able to operate with power dissipation below 1~mW but $\beta$ for these designs cannot be calculated as $OP_1$ measurements are not available. Conversely, the KTWPAs detailed in this work have demonstrated $OP_1 \geq -42$~dBm \cite{howe2025kinetic} with dissipation on the order of 200~nW. This yields $\beta_\text{KTWPA} \sim 0.3$, which is more than an order of magnitude higher than HEMTs \cite{lnf}. Furthermore, reduction of the KTWPA insertion loss \cite{faramarzi2024kinetic} and on-chip integration of lossless dc and pump injection circuits \cite{howe2025compact} are feasible improvements that can drive $\beta_\text{KTWPA}$ towards unity. These results indicate that, from an amplifier-added-noise perspective, as well as in relation to the amplifier efficiency metric and total power dissipation, KTWPAs are a promising alternative to HEMTs. Demonstrating a readout chain based solely on KTWPAs at the cryogenic stages, e.g. the QARC, is intended to expand the parameter space of what is feasible for cryogenic readout chains in highly-scaled or cooling-power-starved applications.

\subsection{KTWPA Principle of Operation}
A KTWPA uses the nonlinear kinetic inductance of a disordered superconductor, which depends on the current in the film
\begin{equation}\label{eq:kin1}
    \mathcal{L}_k(I) = \mathcal{L}_0\left[ 1 + I_p(\varepsilon + \xi I_p) + \mathcal{O}(I^3)\right],
\end{equation}
to realize the required reactive nonlinearity for parametric amplification. The prefactors $\varepsilon$ and $\xi$ respectively describe the relative strength of the three-wave-mixing (3WM) and four-wave-mixing processes (4WM):
\begin{equation}\label{eq:eps}
    \varepsilon = \frac{2 I_{dc}}{I_*^2 + I_{dc}^2}~, ~~~ \xi = \frac{1}{I_*^2 + I_{dc}^2}
\end{equation}
and demonstrate that a predominantly 3WM or 4WM film may be realized with the application or omission of dc bias current.

In the perfect-phase-matched limit the current amplitude at position $x$ is exponential \cite{boyd2008nonlinear}:
\begin{equation}
    I_S = \cosh(g_3 x)I_{S0}
\end{equation}
and thus so is the gain
\begin{equation}\label{eq:gain_analytic}
    G(x) = \left| \frac{I_s(x)}{I_{s0}} \right|^2 = \cosh^2(g_3 x).
\end{equation}
Here $g_3 = \varepsilon I_{p0} \sqrt{k_S k_I} / 4$ is the 3WM element with $I_{p0}$ the initial pump amplitude and $k_S~(k_I)$ the signal (idler) wavenumber. Broadband gain in our KTWPAs is achieved via periodic, stepped-impedance modulation to place a photonic bandgap above the target signal band \cite{howe2025kinetic, giachero2024kinetic}.

\subsection{Experimental Setup and Basic Characterization}
The two amplification chains we study are shown in Fig.~\ref{fig:double_kit_setup} -- namely a double-KTWPA frontend operated at 0.05~K both with (semi-QARC) and without a HEMT (QARC) at 4.5~K. In both cases we use a warm semiconductor amplifier (WAMP) with a gain of 39~dB to ease the burden of signal acquisition and processing. The WAMP is not essential for the amplification chain to reach nQL noise levels but does greatly reduce integration time during noise measurements as it facilitates overcoming the noise temperature of the spectrum analyzer. Characterization of each system is performed in two separate cooldowns. The semi-QARC is not the main product of this work but serves as a useful intermediate stage in bringing up the QARC.

Each KTWPA is wired with its own directional coupler and pair of bias tees to enable fully independent optimization of each KTPWA's tuning point. Diplexers are located at the output of each KTWPA to remove the residual pump tone and prevent compression of the next stage amplifier (KTWPA2 or the HEMT). The directional couplers, bias tees, and diplexers are connectorized, commercially-available microwave components. A known white noise power is generated using a shot-noise tunnel junction (SNTJ) at the input of KTWPA1 -- which permits measurement of the system noise (semi-QARC or QARC) at the KTWPA1 input reference plane \cite{howe2025kinetic, howe2025compact, malnou2024low}. A full experimental schematic is shown in Fig.~\ref{fig:full_schematic}.

\begin{figure*}[t]
    \centering
    \includegraphics[width=0.98\textwidth]{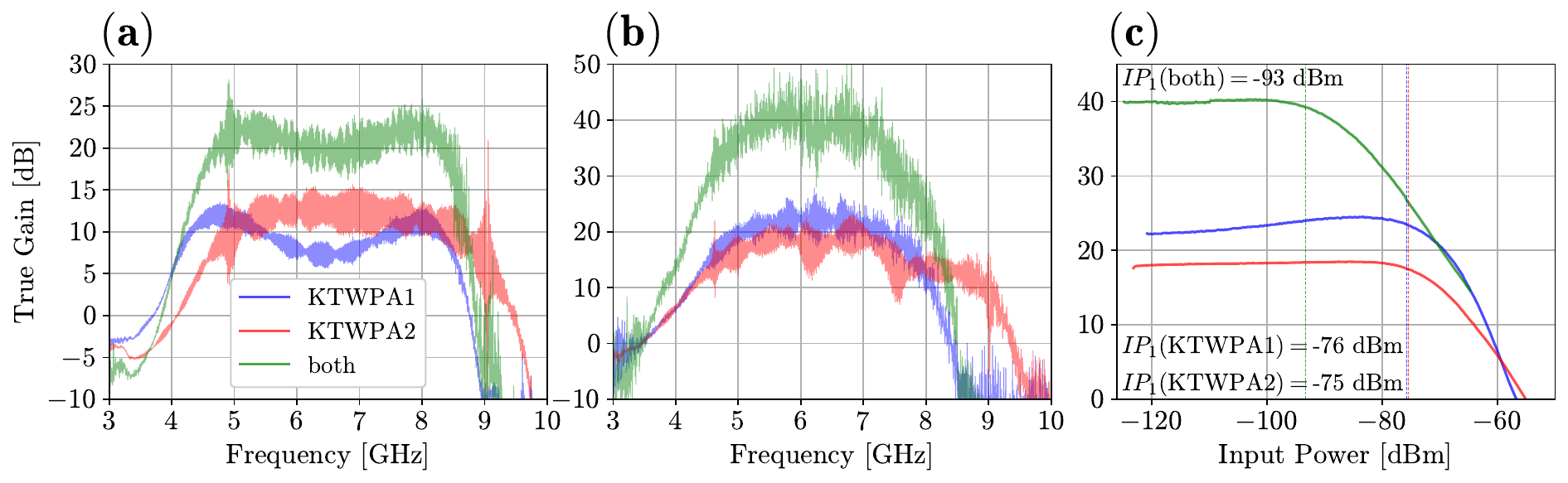}
    \caption{Amplifier chain basic performance metrics as characterized using a VNA. Gain measurements presented here focus on the dual-KTWPA portion of the chain and omit the HEMT and WAMP gains. The VNA is used to measure the pump-on to pump-off ratio to extract the \textit{on/off} gain. The true gain is then computed using insertion loss measurements of each KTWPA, as well as the whole component chain obtained in separate cooldowns using cryogenic rf switches and a matched through cable reference. \textbf{(a)} Semi-QARC separate and in-tandem true gain. Each KTWPA is tuned in this case to yield the largest possible total-chain gain-bandwidth-product (GBP); resulting in the bandwidth-optimized nQL operating point. \textbf{(b)} QARC gain optimized to produce maximum gain rather than GBP. This procedure is necessary to overwhelm the WAMP-added-noise over an appreciable bandwidth and realize nQL system noise using the QARC -- which is more difficult without a HEMT. Example QARC 1~dB input compression ($IP_1$) power measurement at 6.5~GHz when operated at maximum gain as in (b). This is the minimum $IP_1$ achievable by the QARC and will increase should the QARC be operated at a GBP-optimized tuning, as in (a), which would reduce the maximum in-tandem KTWPA gain.}
    \label{fig:qarc_semi_qarc_gain_IP1}
\end{figure*}

Implementation of a second KTWPA does require additional pump and bias lines and as well as a larger experimental footprint due to numerous bulky, external commercial rf components. This can hamper widespread realization of such multi-stage systems, however, successful on-chip integration for a single device has been demonstrated \cite{howe2025compact}. While dissipation of the pump is currently dominated by the insertion loss of the KTWPA and not the external pump injection circuit, a superconducting bias tee, with bias supplied by superconducting wiring, eliminates the $\sim 100$~nW of dc dissipation -- which effectively doubles the scalability of such a system from a dissipation perspective. In terms of experimental volume, the total footprint of a KTWPA with integrated pump and bias injection can be less than one-fifth that of a setup using external commercial components.

Basic gain and compression measurements for the amplifier chains studied in this work are shown in Fig.~\ref{fig:qarc_semi_qarc_gain_IP1}; where we report only the KTPWA gains. Due to the fact that less KTWPA gain is necessary to set the system noise to the intrinsic KTWPA limit in the semi-QARC case (discussed further in the following section) we tune both KTPWAs to maximize the gain bandwidth product (GBP). This technique, shown in Fig.~\ref{fig:qarc_vs_semi_qarc_noise}(a), allows us to demonstrate the widest nQL noise performance possible in the following section. Conversely, for Fig.~\ref{fig:qarc_vs_semi_qarc_noise}(b) the KTWPAs are tuned to maximize their total gain such that we reach the minimum $\sim 40$~dB of parametric gain necessary to overwhelm the WAMP's added noise at room temperature.

\section{Noise Model}
The noise model can be derived by considering each stage of the amplifier chain \cite{malnou2021three, malnou2024low}. In photon-normalized units the output power of a phase-insensitive parametric amplifier is

\begin{equation}
    N_\text{1,out}^{S} = G_1^{S} \left( N_\text{1,in}^{S} + \frac{G_1^{I_1}}{G_1^{S}} N_\text{1,in}^{I_1} + N_\text{1,ex}^{S} \right)
\end{equation}
where $G_n$ is the gain of the $n$-th parametric amplifier stage, subscripts and superscripts indicate evaluation at the $n$-th stage. The \mbox{\{in, out\}} subscripts denote whether the term describes the noise at the input or output of the $n$-th stage. For both amplifiers the signal is the same and only the idler changes due to each stage being pumped at a different frequency. $N_\text{1, ex}^S$ encapsulates the noise in excess of the QL. In practice the signal frequencies for each KTWPA stage are the same but the idlers need not be -- which occurs when the two amplifiers are not pumped at identical frequencies (our case). Thus the output of the second KTWPA is
\begin{equation}
    N_\text{2,out}^{S} = G_2^{S} \left( N_\text{1,out}^{S} + \frac{G_2^{I_2}}{G_2^{S}} N_\text{1,out}^{I_2} + N_\text{2,ex}^{S} \right)
\end{equation}
\begin{equation}
\begin{split}
    N_\text{2,out}^{S} = ~& G_2^{S}\left[ G_1^{S} \left( N_\text{in}^{S} + \frac{G_1^{I_1}}{G_1^{S}} N_\text{in}^{I_1} + N_\text{1,ex}^{S} \right) + \right. \\  
    & ~~~~~~~~~~~~~~~~~~~~~ \left. \frac{G_2^{I_2}}{G_2^{S} }N_\text{1,out}^{I_2} + N_\text{2,ex}^{S} \right].
\end{split}
\end{equation}
\begin{equation}
\begin{split}
    N_\text{2,out}^{S} = ~& G_1^{S} G_2^{S} \left( N_\text{in}^{S} + \frac{G_1^{I_1}}{G_1^{S}} N_\text{in}^{I_1} + N_\text{1,ex}^{S} \right) + \\
    & G_2^S \left( \frac{G_2^{I_2}}{G_2^{S} }N_\text{1,out}^{I_2} + N_\text{2,ex}^{S} \right)
\end{split}
\end{equation}
As long as $G_1^S \gg 1$ a multi-stage parametric amplifier chain does not incur a penalty via the existence of multiple idlers (one per quantum amplifier which requires a distinct pump) contributing 0.5~quanta each.

The third stage output is
\begin{equation}
    N_\text{3,out}^S = G_3(N_\text{2,out}^S + N_\text{3,add}^S)
\end{equation}
\begin{equation}
\begin{split}
    N_\text{3,out}^{S} = ~& G_1^{S} G_2^{S} G_3^S \left( N_\text{in}^{S} + \frac{G_1^{I_1}}{G_1^{S}} N_\text{in}^{I_1} + N_\text{1,ex}^{S} \right) + \\
    & G_2^S G_3^S \left( \frac{G_2^{I_2}}{G_2^{S} }N_\text{1,out}^{I_2} + N_\text{2,ex}^{S} \right) + G_3^S N_\text{3,add}^S
    \label{eq:Nout_full_chain}
\end{split}
\end{equation}
because in both the semi-QARC and QARC studied here the third stage is a semiconductor amplifier -- where $N_\text{3,add}^S$ is its added noise. In the semi-QARC case, as long as $N_\text{3,add}^S$ does not exceed $\sim 20$~quanta, $G_1^S$ and $G_2^S$ of only 10~dB is sufficient to set the chain noise at nQL levels. For the semi-QARC this is a simple feat as state-of-the-art HEMTs at 4~K supply added noise of around 30~quanta or less. However, for the QARC the third stage is a gain block semiconductor amplifier with a 3~dB noise figure (the WAMP) -- indicating $N_\text{3,add}^S$ is approximately 300~K ($\sim 1250$~quanta) when its input sees a room temperature load. The break-even total QARC gain is 34~dB in this case.

The high-first-stage-amplifier gain limit reduces Eq.~\eqref{eq:Nout_full_chain} to
\begin{equation}
    N_\text{chain,out}^S = G_\text{sys}^S \left( N_\text{1,in}^S + \frac{G_1^{I_1}}{G_1^S}N_\text{1,in}^{I_1} + N_\text{1,ex}^S \right)
    \label{eq:chain_input_output}
\end{equation}
$N_\text{1,ex}^S$ quantifies the degree to which a parametric amplifier fails to meet the QL of added noise -- or to set the amplifier chain added noise to the QL as the second term in parentheses embodies the QL. Any failure of the first-stage-amplifier to set the system noise to its intrinsic limit -- due to later stage added noise or insufficient gain -- is allocated to the first stage's excess noise and thus provides an upper bound for the TWPA's intrinsic added noise. 

\section{QARC and Semi-QARC System Noise Measurements}
\subsection{Experimental Protocol}
To measure the amplifier chain system noise one must establish a relation between chain input and output noise and the $y$-intercept of the resulting fit to Eq.~\eqref{eq:chain_input_output} yields the excess noise. To this end we use an intrinsically calibrated noise source, realized using a shot noise tunnel junction (SNTJ), to measure the amplifier chain system noise \cite{howe2025kinetic, malnou2021three}. The SNTJ noise power spectral density in photon-normalized units is
\begin{align} 
    N_\text{SNTJ} = & \frac{eV + \hbar \omega}{4 \hbar \omega} \coth\left( \frac{eV + \hbar \omega}{2 k_B T_e} \right) + \\
    & \frac{eV - \hbar \omega}{4 \hbar \omega} \coth\left( \frac{eV - \hbar \omega}{2 k_B T_e} \right),
    \label{eq:sntj_psd}
\end{align}
with $T_e$ the electron temperature, $\omega$ the signal frequency, and $e,~\hbar,~k_B$ the electron charge, reduced Planck constant, and Stefan-Boltzmann constant, respectively. Operating the SNTJ in the \textit{high-voltage-bias limit} \cite{malnou2024low, howe2025kinetic}gives a linear relation between the voltage bias across the SNTJ and the generated noise:
\begin{equation}
    N_\text{SNTJ} \rightarrow \frac{e |V|}{2\,\hbar\,\omega}.
\end{equation}
This limit is useful as it provides a linear relation between the bias and generated noise which is independent of $T_e$ and robust to experimental nonidealities \cite{malnou2024low} -- i.e. accessing the high-bias limit portion of the curve provides the most stable and accurate measurements and analysis.

During measurement the SNTJ is biased with a 10~Hz triangle wave from an arbitrary waveform generator (AWG) and the chain-output noise of Eq.~\eqref{eq:chain_input_output} is captured on a spectrum analyzer (SA). The SA is operated in zero-span mode and triggered by the AWG with a sweep time equal to one AWG period. The SNTJ impedance are determined \textit{in situ} using four-wire resistance measurements to accurately transfer the AWG bias through the $\sim 10$~k$\Omega$ bias resistors and obtain accurate calibration of the SNTJ voltage bias. For the device used in this work the impedance is $Z_\text{SNTJ} = 41.8~\Omega$.

In this way, the power input to the amplifier chain at the signal frequency is $N_\text{1,in}^S = N_\text{SNTJ}^S$. The SA captures the full chain output power (i.e. WAMP output power) $N_\text{SA} = N_\text{WAMP,out}$ at a single frequency with a $B = 8$~MHz resolution bandwidth. The $N_\text{SA} = P_\text{SA} / (h\,f\,B)$ trace linearly maps the chain output power to SNTJ voltage and thus, input noise. To determine the mapping between SNTJ voltage and SA trace point we perform a fit to 
\begin{equation}
    N_\text{chain,out}^S = G_\text{sys}^S N_\text{SNTJ},
\end{equation}
where in $N_\text{SNTJ}$ we allow parasitic voltage offsets via $V \rightarrow V - V_\text{off}$, to properly align the SA and AWG sampling. This procedure establishes the requisite voltage-to-noise relation.

Next, we address the fact that it is desirable to characterize the system noise at a reference plane where an experimenter plans to locate a device-under-test. To move the reference plane off the SNTJ die we apply the usual beam splitter relation to account for the SNTJ-packaging mismatch and the insertion loss of the interstitial bias tee required to operate it:
\begin{equation}
    N_\text{1,in}^s = \eta N_\text{SNTJ} + (1 - \eta) N_{T_f},
\end{equation}
where $\eta$ is the power transmission efficiency between the SNTJ (on-chip) and the output of the dc block on the other side of its bias tee. This places the noise calibration reference plane at the input of the dual-KTWPA amplifier chain (at the directional coupler input of Fig.~\ref{fig:double_kit_setup}) and represents the closest point an experimenter can locate a readout target device. The SNTJ-package mismatch has been measured previously \cite{howe2025kinetic} (300~K measurement) and is nearly flat with an average value of -0.7~dB at cryogenic temperatures in the frequency range of interest \cite{howe2025kinetic} -- resulting in $\eta = 0.8$ (see Appendix~\ref{sec:error_analysis}).

Finally, we select the high-bias portion of the noise curve, fit Eq.~\eqref{eq:chain_input_output}, and extract the system gain and excess noise. This procedure is shown in detail using an alternate noise source, but whose operation is effectively identical to the SNTJ, in the final section of this article.

\subsection{Uncertainty Analysis}
In general, our measurements are characterized by moderately nonstationary variances and mild short-range autocorrelation. Therefore, per-point uncertainties are estimated using a joint two-stage heteroskedasticity-and-autocorrelation-consistent (HAC) sandwich covariance estimator. This is then validated using a circular block-multiplier bootstrap \cite{newey1987simple, kunsch1989jacknife}. Of the free parameters not directly addressed by the statistical treatment, $\eta$ plays the biggest role in determining uncertainty and is propagated separately. Details of the uncertainty analysis are provided in Appendix~\ref{sec:error_analysis}.

\begin{figure}[t]
    \centering
    \includegraphics[width=0.48\textwidth]{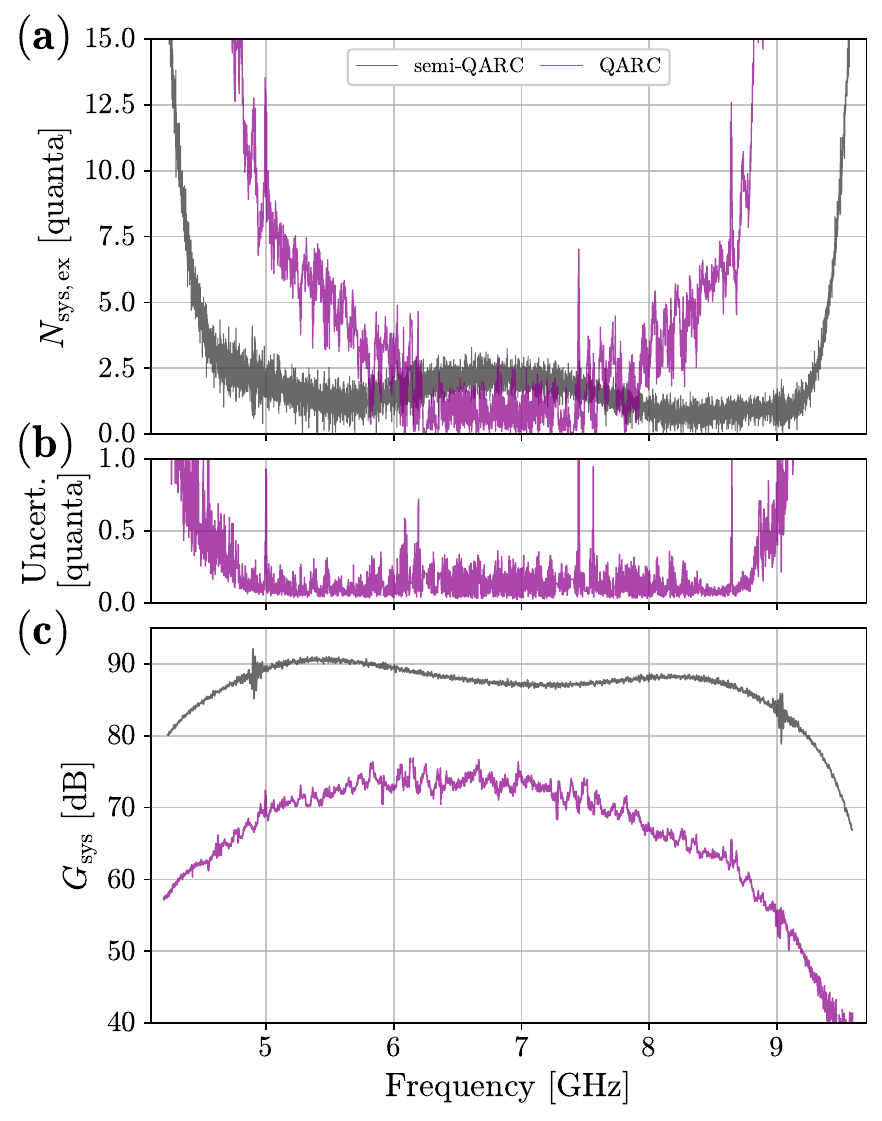}
    \caption{SNTJ-assisted noise measurement of the semi-QARC and QARC. \textbf{(a)} System excess noise, $N_\text{sys,ex}$. \textbf{(b)} Measurement uncertainty in $N_\text{sys, ex}$ for the QARC using a HAC sandwich estimator with the $\eta$ errorbar propagated independently. For brevity we show only the QARC case as the semi-QARC case displays nearly identical behavior. \textbf{(c)} Extracted system gain, $G_\text{sys}$. Note that the QARC gain is not exactly the semi-QARC gain minus the HEMT gain due to the different choices in KTPWA tunings -- i.e. the semi-QARC case maximizes the bandwidth where the chain operates below 2~quanta of excess noise, while the QARC requires pure maximum KTWPA gain optimization to reach similar excess noise.}
    \label{fig:qarc_vs_semi_qarc_noise}
\end{figure}

\subsection{Results}
Initially, we measure the system noise of the semi-QARC to establish a baseline for the expected system noise using the double-KTWPA frontend as this is a minimal perturbation of amplification chains typically encountered in nQL cryogenic systems \cite{krantz2019quantum}. Indeed, we should expect similar ultimate system noise to previously demonstrated semi-QARC systems -- limited by the intrinsic added noise of our KTWPAs \cite{howe2025kinetic, malnou2021three, faramarzi2024kinetic, ranadive2024traveling, macklin2015near}. However, it is easier in a double-KTWPA frontend architecture for the first stage amplifier (KTWPA1) to dominate the second stage amplifier's added noise because this is now another KTPWA rather than a HEMT -- with an added noise more than a factor of ten lower than state-of-the-art HEMTs \cite{malnou2024low}.

Results of the noise measurements for the semi-QARC and QARC -- both tuned according to Fig.~\ref{fig:qarc_semi_qarc_gain_IP1} -- are shown in Fig.~\ref{fig:qarc_vs_semi_qarc_noise}(a). Fig.~\ref{fig:qarc_vs_semi_qarc_noise}(b) shows the extracted system gains for each amplifier chain and tuning. The semi-QARC tuning demonstrates nearly 5~GHz of bandwidth where the system excess noise is below 5~quanta. With the QARC we obtain nearly 2~GHz of bandwidth where the mean excess noise is also 5~quanta or less. This is the first time an amplifier chain omitting a cryogenic semiconductor amplifier has operated this near to the QL over these bandwidths. 

For the QARC we observe that the total KTWPA gain is only barely sufficient to overwhelm the WAMP added noise over much of its bandwidth. This is evident in the reduced bandwidth, elevated ultimate noise, and larger sensitivity in $N_\text{sys,ex}$ to the gain ripple. In the semi-QARC the gain at each stage is sufficient to overwhelm the next stage's added noise, from which we can conclude that the ripple in $N_\text{sys,ex}$ for the semi-QARC is due to intermodulation distortion or spurious frequency conversion. For the QARC the gain ripple has an additional effect wherein the double-KTPWA frontend fails at some frequencies -- at the gain ripple nadirs away from the central high gain region -- to dominate the WAMP added noise. This effect dominates that of additional loss incurred from bias tees, directional couplers, and the additional isolator which replaces the HEMT. The total insertion loss of these components is on the order of 2~dB and the gain ripple is sub-dominant to the 8~dB to 10~dB QARC gain ripple. Nonetheless, the flattening of the QARC excess noise between approximately 6~GHz and 7.5~GHz to the same ultimate noise level as the semi-QARC demonstrates a wide bandwidth over which the dual-KTWPA frontend indeed overwhelms the WAMP-added noise and for which the amplifier chain operates at the KTWPA added noise intrinsic limit of $N_\text{sys,ex}^\text{ult} = 0.8 \pm 0.7$~quanta.

\begin{figure*}[t]
    \centering
    \includegraphics[width=0.98\textwidth]{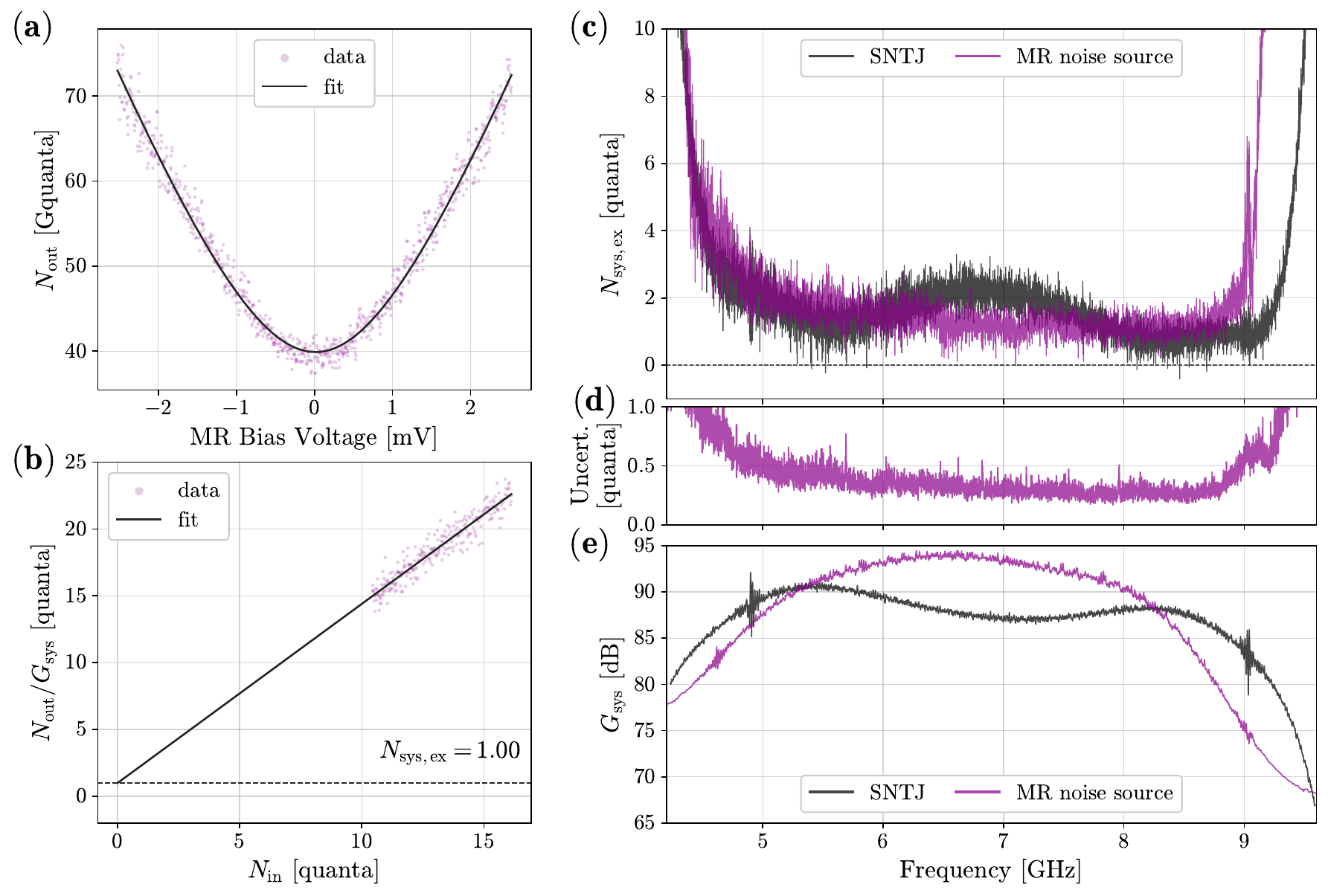}
    \caption{Operation of the MR noise source and semi-QARC noise measurement using both the SNTJ and MR noise source. \textbf{(a)} Semi-QARC output power as a function of the applied MR bias voltage with the spectrum analyzer frequency set to 7~GHz. The curve $G_\text{sys} N_\text{HE}$ is fit to extract the system gain and align the SA trace to the AWG voltage ramp. This process provides the mapping between MR voltage, $N_\text{in}$, and the SA data. \textbf{(b)} Linear fit of Eq.~\eqref{eq:chain_input_output} to the high-bias portion of (a). The $y$-intercept  yields the semi-QARC excess noise of 1.0~quanta at 7~GHz with this tuning. \textbf{(c)} Full system excess noise, $N_\text{sys,ex} = N_\text{1,ex} + N_\text{2,ex}$, frequency sweep using first an SNTJ and then a MR noise source in separate cooldowns and with different semi-QARC tunings. In the MR case the two KTWPAs are tuned to produce the symmetric, maximum peak gain profiles of the QARC tuning presented in the previous section. The SNTJ data is the same as is shown in Fig.~\ref{fig:qarc_vs_semi_qarc_noise}(a) and corresponds to the semi-QARC tuning which maximizes the GBP. \textbf{(d)} Semi-QARC system noise uncertainty using the mesoscopic resistor noise source. \textbf{(e)} Extracted system gain for each measurement.}
    \label{fig:Rmeso_noise_master}
\end{figure*}

\section{Validation of a Diffusive Nanowire Noise Source}
In the final section of this article we validate a noise source which can serve as an alternative to an SNTJ. SNTJs are often constructed from Josephson junctions whose superconductivity is suppressed by strong rare earth magnets placed in proximity to the junction \cite{chang2016noise, malnou2024low}. Josephson junction magnetic field susceptibility makes integration of these SNTJs with quantum information \cite{krantz2019quantum, rower2023evolution} and single-flux-quantum devices \cite{mukhanov2019josephson} challenging. Further, in strong ambient field applications, such as axion searches, this prevents hardware configurations which woudl directly enable in situ measurement of the experiment's sensitivity (added noise) using SNTJs due to strong magnetic forces between the SNTJ magnet and axion conversion field. From a metrological perspective, low-pass filters formed from the intrinsic junction inductance, capacitance, and 50~$\Omega$ impedance impose low frequency cutoffs in the white shot noise spectrum on the order of 30~GHz to 50~GHz. Above these rolloff frequencies the primary utility of the SNTJ -- the fact its output noise power can be known accurately via only a single, simple-to-measure parameter -- is spoiled. In principle this critical frequency is straightforward to measure and the super-cutoff response may be ascertained, but this is likely to vary significantly on a device-level basis and undermines the \textit{intrinsically-calibrated} nature of these noise sources. As a final practical note, SNTJs for microwave noise calibration are often made of small junctions with quite thin barriers and are correspondingly prone to damage via electrostatic discharge.

Instead, for applications using sensitive quantum systems, strong magnetic fields (axion searches), and high frequency ($\geq 20$~GHz) parametric amplifier development, an alternative magnet-free noise source with similar ease-of-use and intrinsic-calibration to that of the SNTJ is attractive. A metallic, diffusive resistor whose length is shorter than the electron-phonon scattering length (a \textit{diffusive nanowire}) can be used to generate self-calibrated noise in an analogous fashion to an SNTJ \cite{shen2006low, teufel2008superconducting, bergeal2012mesoscopic}. Such a mesoscopic resistor (MR) can in principle operate in two regimes. First, the hot electron regime, where the nanowire length lies below the electron-phonon scattering length but above the electron mean free path. Second, in the shot-noise regime, where the length is below the electron mean free path. From a practical perspective the shot-noise regime is difficult to access while simultaneously realizing an impedance-matched (50~$\Omega$) MR as this tends to push MR lengths to below 1~$\mu$m and widths below 100~nm \cite{nagaev1995influence, steinbach1996observation}. However, the hot electron regime is a possible solution as the MR lengths are approximately 3~$\mu$m for Au and Cu. Further, the hot electron spectrum is white up to frequencies on the order of the waveguide cutoff frequency for both the on-chip and off-chip transmission line interconnects for integrating subsequent components (amplifiers, detectors, etc.). Finally, we note that such devices have previously been used in bolometric development \cite{shen2006low, teufel2008superconducting} and quantum amplifier added-noise characterization \cite{bergeal2012mesoscopic}. 

The noise power spectral density emitted by the MR noise source at angular frequency $\omega$ is determined by the diffusive nanowire potential difference via the nanowire effective temperature $T_\text{eff}$ \cite{bergeal2012mesoscopic}:
\begin{equation}
    N_\text{HE}(\omega) = \frac{1}{2} \coth \left( \frac{\hbar \omega}{2 k_B T_\text{eff}} \right).
    \label{eq:mr_psd}
\end{equation}
Here $N_\text{HE}$ is in photon-normalized units, $\hbar$ is the reduced Planck constant, $k_B$ is the Boltzmann constant, and the effective temperature is given by
\begin{equation}
    T_\text{eff} = \frac{T_0}{2} \left[ 1 + \left( \zeta \frac{V}{T_0} + \frac{1}{\zeta} \frac{T_0}{V} \right) \tan^{-1}\left( \zeta \frac{V}{T_0} \right) \right].
    \label{eq:mr_Teff}
\end{equation}
$T_0$ is the temperature of the MR electronic reservoirs (500~nm thick copper rf circuit), $\zeta = \sqrt{3} e / 2 \pi k_B$, and $V$ is the nanowire voltage.

Similar to the SNTJ, the MR has an effective high bias limit when $\zeta \gg 1$. in photon-normalized units
\begin{equation}
    N_\text{HE} \rightarrow \frac{\sqrt{3} \,e\,|V|}{8\,h\,f},\qquad T_\text{eff} \rightarrow \frac{\sqrt{3} \,e\,|V|}{8\, k_B}
\end{equation}

The MR noise source does have an additional free parameter ($T_0$) when compared to the SNTJ however, just as in the SNTJ case, the high bias limit is independent of this factor -- making the MR noise source an effective SNTJ-alternative. Here we perform initial measurements of the QARC system noise extracted using the hot electron source fabricated from a 25~nm evaporated copper film with a 100~nm width and 3~$\mu$m length. The \textit{bath} for the nanowire is composed of the 500~nm thick Cu rf circuit everywhere but where the nanowire resides. We compare noise measurements of the semi-QARC using the MR source to the SNTJ-assisted measurements of the previous section to cross-validate the MR. These measurements are performed in two sequential cooldowns in which the only change to the experimental setup of Fig.~\ref{fig:double_kit_setup} is to replace the SNTJ package with the MR package.

The measurement protocol using the MR noise source is identical to the SNTJ described above -- with the only difference being the form of the calibrated noise being generated \cite{howe2025kinetic, howe2025compact, malnou2024low, malnou2021three}. Explicitly, we again consider the high-gain limit for KTWPA1, which is Eq.~\eqref{eq:chain_input_output}, with $N_\text{1,in}^{S,I_1} = N_\text{HE}^{S,I_1}$. We also measure the MR impedance in the same manner and obtain $Z_\text{MR} = 41.2~\Omega$ for the device used herein.

The semi-QARC MR noise measurement results and comparison with the SNTJ case are detailed in Fig.~\ref{fig:Rmeso_noise_master}. In Fig.~\ref{fig:Rmeso_noise_master}(a) we show an example of capturing the generated noise of the MR as a function of its bias voltage on a spectrum analyzer at 6~GHz. Fig.~\ref{fig:Rmeso_noise_master}(b) establishes the expected linear relationship between $N_\text{in}$ and $N_\text{out}$ where we use the linear high bias regime ($V_\text{MR} \gtrsim 1$~mV). In Fig.~\ref{fig:Rmeso_noise_master}(c) and (d) we show the full spectrum analyzer frequency sweep where the system excess noise and system gain is constructed via the procedure shown in Fig.~\ref{fig:Rmeso_noise_master}(a) and (b) at every frequency. We also show the same SNTJ-assisted semi-QARC noise measurement shown in Fig.~\ref{fig:qarc_vs_semi_qarc_noise}(a). Despite different tunings for the two KTWPAs, both measurements with the two noise sources show excellent agreement in the ultimate system noise and are obtained with almost no modification in the experimental workflow for collecting an amplifier noise measurement. Finally, Fig.~\ref{fig:Rmeso_noise_master}(d) and (e) show the QARC excess noise uncertainty and system gain, respectively.

\section{Conclusion}
Scaled cryogenic quantum information systems, space-based observatories and other high-device-density applications require low-noise and low-dissipation amplifiers. Here we have presented a direct comparison of an amplification chain based on a double-stage KTWPA frontend. We configure the chain both with (semi-QARC) and without (QARC) a follow-on semiconductor HEMT amplifier at 4.5~K. With the semi-QARC tuned for the maximum double-KTWPA gain bandwidth product we demonstrate nearly 5~GHz of bandwidth where the readout chain system excess noise is below 2~quanta. We then compare this to the QARC system noise by replacing the HEMT with an isolator and operate the KTWPAs in their gain-maximizing mode. With this tuning the QARC achieves a competitive system excess noise with the semi-QARC over the QARC's $\sim$2~GHz 3~dB bandwidth -- as well as a narrower $\sim$1.2~GHz region where the QARC excess noise reaches the KTWPA intrinsic limit of $\sim 1.3$~quanta. These results show nQL performance is accessible in readout chains without semiconductor follow-on amplifiers inside the cryostat. Further, omission of semiconductor amplifiers reduces the burden on the cryogenic cooling system by orders of magnitude and improves the scalability and orbital-feasibility of nQL readout chains.

Additionally, we use the semi-QARC to cross-validate a newly fabricated noise source based on mesoscopic diffusive nanowires which create self-calibrated hot electron noise. We measure the semi-QARC system noise using an SNTJ and the MR source and our preliminary analysis shows that the two noise sources provide consistent results over the semi-QARC bandwidth. This demonstrates that mesoscopic nanowires are a promising tool for measuring amplifier chain added noise with potential application in spaces where SNTJs are less optimal or entirely prohibited.

\begin{acknowledgments}
This work is supported by the NIST Innovations in Measurement Science program, the National Aeronautics and Space Administration (NASA) under Grant No. NNH18ZDA001N-APRA, the Department of Energy (DOE) Accelerator and Detector Research Program under Grant No. 89243020SSC000058, and DARTWARS, a project funded by the European Union’s H2020-MSCA under Grant No. 101027746 A.~Giachero is supported by the Italian National Quantum Science and Technology Institute through the PNRR MUR Project under Grant PE0000023-NQSTI.

We thank M. Malnou, and J. Aumentado of the NIST Advanced Microwave Photonics Group for providing the SNTJ used in this work.
\end{acknowledgments}

\appendix

\section{Complete Experimental Schematic}
\begin{figure*}
    \centering
    \includegraphics[width = \textwidth]{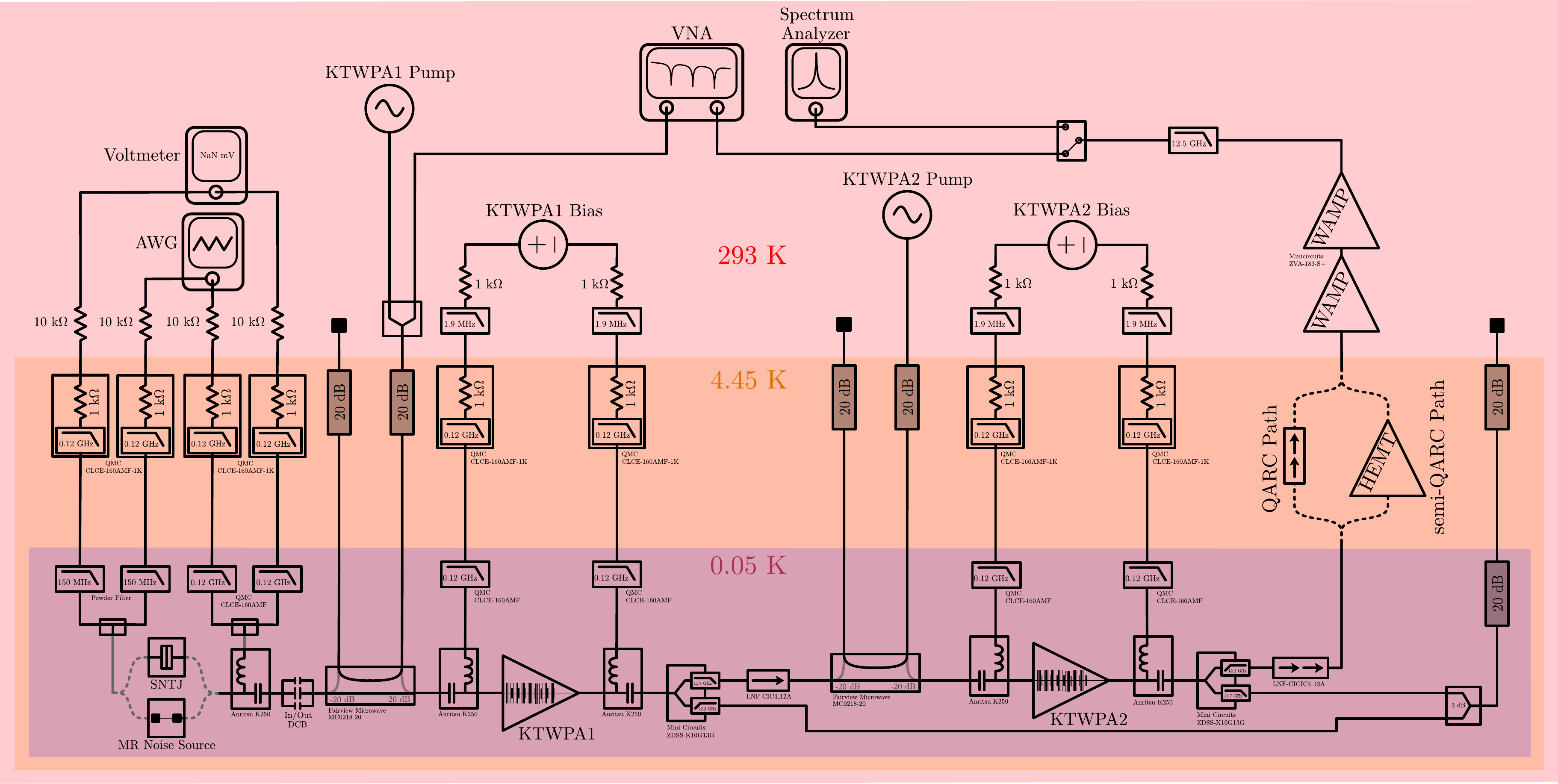}
    \caption{Complete experimental schematic for all configurations and measurements detailed in the main body. All connections both outside and inside the cryostat are made using coaxial cables; including dc bias lines. Connections indicated in gray (SNTJ and MR noise source bias and voltage leads) are made using cables where continuity in the outer cable shield is broken to permit differential biasing of the SNTJ. Diplexers at the output of each KTWPA are used to split off the signal and idler bands from the pump to aovid compression of the updtream amplifiers. These are joined using a power combiner and dumped/attenuated on a separate coaxial line (far right).}
    \label{fig:full_schematic}
\end{figure*}

Fig.~\ref{fig:full_schematic} shows the complete experimental schematic for all measurements used in the main text \footnote{Commercial instruments and software are identified in this paper in order to adequately specify the experimental procedure. Such identification does not imply recommendation or endorsement by NIST, nor does it imply that the product identified is necessarily the best available for the purpose.}.

\section{System Excess Noise Error Analysis}
\label{sec:error_analysis}
Here we describe our framework for extracting the uncertainty estimator for the system excess noise. The procedure is the same for the SNTJ-QARC measurement shown in Fig.~\ref{fig:qarc_vs_semi_qarc_noise}(b) and for the MR-semi-QARC of Fig.~\ref{fig:Rmeso_noise_master}(d). As described in the main text, at each frequency, the calculation consists of the first-stage calibration fit to either the SNTJ or MR theoretical output noise -- Eqs.~\eqref{eq:sntj_psd} or \eqref{eq:mr_psd} -- to create the mapping between noise source voltage and the SA trace and extract $G_\text{sys}$. This is then followed by identification of the high-bias regime and fitting to Eq.~\eqref{eq:chain_input_output} to extract $N_\text{sys,ex}$.

Explicitly, in the first stage, the SA-captured bias sweep is fitted to the noise source output power,
\begin{equation}
    Y_i = G_{\mathrm{sys}}
          N^{\mathrm{SNTJ,HE}}_i(V_i, T_{0,e}, V_0) + \epsilon_i,
    \label{eq:rmeso-first-stage}
\end{equation}
where $Y_i$ is the measured output-noise power, $G_{\mathrm{sys}}$ is the system gain, $T_0$ is the fitted zero-bias electron temperature, and $V_0$ is the voltage-SA trace offset. $V_0$ is treated as a fixed parameter for a given added noise frequency sweep so extraction is performed only once when evaluating the source model. Above, $\epsilon_i$ denotes the regression error for the $i$-th sample

The second stage fit is performed to the positive-branch high-bias samples with $V_{\mathrm{MR}}\geq 1.0\,\mathrm{mV}$. The first-stage gain was fixed, and the identifiable two-parameter relation
\begin{equation}
    \frac{Y_i}{G_{\mathrm{sys}}}
      = a N_{\mathrm{in},i} + N_{\mathrm{sys,ex}}
    \label{eq:rmeso-downstream}
\end{equation}
was fitted directly to the raw samples.  Here $a$ is a nuisance slope
that absorbs relative gain or transmission errors but is distinct from uncertainty in the noise source-amplifier-to-chain-input power transmission $\eta$. The noise incident on the amplifier was modeled as
\begin{equation}
    N_{\mathrm{in},i}
      = \eta N_i^{\mathrm{SNTJ,HE}}
        + (1-\eta)\frac{1}{2},
    \label{eq:rmeso-input-noise}
\end{equation}
with nominal power transmission $\eta_1=0.7$.  Fixing
$G_{\mathrm{sys}}$ to its independently determined first-stage value
avoids the degeneracy that would result from fitting the downstream
gain, slope, and intercept simultaneously.

The statistical uncertainty is obtained from the joint estimating
equations for the three first-stage parameters
$(G_{\mathrm{sys}},T_{e,0},V_0)$ and the two downstream parameters
$(a,N_{\mathrm{sys,ex}})$.  A heteroskedasticity- and
autocorrelation-consistent sandwich covariance matrix was calculated as
\begin{equation}
    \widehat{\mathbf{C}}
      = \mathbf{B}^{-1}\mathbf{M}
        \left(\mathbf{B}^{-1}\right)^{\mathsf{T}},
    \label{eq:rmeso-sandwich}
\end{equation}
where $\mathbf{B}$ is the derivative (``bread'') matrix of the joint
estimating equations and $\mathbf{M}$ is a Bartlett-kernel estimate of
the long-run covariance of the bias-ordered score contributions. A baseline correlation length of $L=10$ samples is used, approximately $n^{1/3}$ for the 901-point first-stage SNTJ or MR sweep.  Because the same bias samples contribute to both fits, this joint construction retains the covariance between the first-stage calibration and downstream intercept.  The pointwise statistical standard uncertainty in the excess noise is then extracted from the corresponding diagonal entry of the HAC covariance matrix
\begin{equation}
    u_{\mathrm{stat}}
      = \sqrt{
          \widehat{C}_{N_{\mathrm{sys,ex}},N_{\mathrm{sys,ex}}}
        }.
    \label{eq:rmeso-statistical-uncertainty}
\end{equation}

We then cross-validate the HAC sandwich approximation was validated using a nonlinear block-multiplier bootstrap technique using contiguous circular bias blocks. Independent mean-one exponential weights were
assigned blockwise, the same weights were applied to both fitting
stages, and both stages were refitted in every replicate.  The
validation included 5, 6, 7, 8, and $9\,\mathrm{GHz}$; block lengths
$L=5$, 10, and 20; and 1000 replicates per frequency and block length,
for 15000 complete refits. All refits converged. The largest
difference between the sandwich and bootstrap standard uncertainties
was $6.49\%$, and the maximum bootstrap bias was $0.111$ standard
deviations. Bootstrap bias was therefore retained only as a validation
diagnostic; it was neither used to correct the reported estimate nor
folded into a root-mean-square error. Variation over $L=5$--20 was
treated as a bandwidth-sensitivity test rather than as an additional
independent uncertainty with consistent results -- thus we conclude the HAC method is a robust statistical uncertainty estimator.

\begin{figure}
    \centering
    \includegraphics[width=0.95\linewidth]{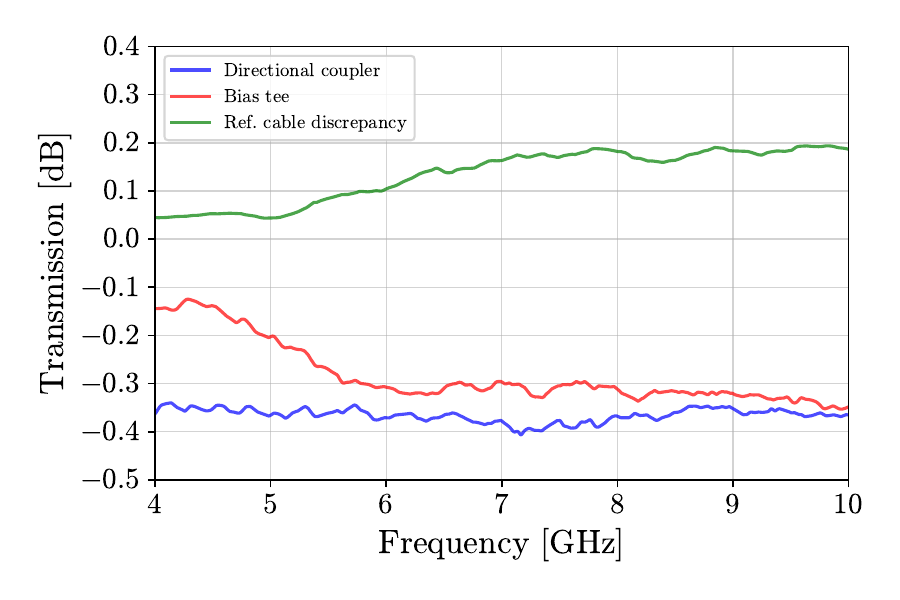}
    \caption{Cryogenic power transmission efficiency of one bias tee and directional coupler part used for KTWPA bias and pump injection. Components are connected to two Sp6T rf switches using matched 10~cm cables. We also connect an SMA barrel in two switch positions -- also using matched cables -- to measure discrepancy between two nominally identical paths. We treat this as our uncertainty in the loss measurements with a characteristic magnitude of $\delta \eta = 0.1$~dB.}
    \label{fig:insertion_losses}
\end{figure}

As it has the largest individual effect on the extracted excess noise, we treat the known uncertainty in $\eta$ as a fixed systematic independent of frequency and propagate it on top of the HAC uncertainty. To do this we configure two Sp6T cryogenic rf switches (0--18~GHz bandwidth) with a bias tee and directional coupler used to inject pump and bias to the KTWPAs on different switch branches. These components, which make up $\eta$, are connected using matched length cables and the third and fourth switch path serve as a cable calibration path with an SMA barrel connector between the cables -- i.e. this approximates a \textit{zero-loss} path. Fig.~\ref{fig:insertion_losses} shows the results of these measurements. For the combined coupler and bias tee we obtain a mean of $\eta = -0.7$~dB and use this value in our noise measurement analysis.

A Gaussian standard uncertainty of $0.1\,\mathrm{dB}$ in power transmission is propagated through the complete downstream estimator using 21-point Gauss--Hermite quadrature. This produced the pointwise calibration contribution $u_{\eta_1}$. Assuming independence between measurement scatter and the independently determined transmission calibration, the reported marginal standard uncertainty is
\begin{equation}
    u_{\mathrm{tot}}(f)
      = \sqrt{
          u_{\mathrm{stat}}^2(f) + u_{\eta_1}^2(f)
        }.
    \label{eq:rmeso-total-uncertainty}
\end{equation}
Although $u_{\eta_1}$ enters this pointwise quadrature sum, it represents
a frequency-correlated calibration uncertainty rather than independent
noise at each frequency.

For example, at $7\,\mathrm{GHz}$, the MR semi-QARC excess noise result is
\begin{equation}
    N_{\mathrm{sys,ex}} = 0.99 \pm 0.35
    \ \text{quanta}.
    \label{eq:rmeso-seven-ghz-result}
\end{equation}
Its components are $u_{\mathrm{stat}}=0.35$ quanta and
$u_{\eta}=0.015$ quanta.  Thus, the $0.1\,\mathrm{dB}$ uncertainty in
$\eta$ contributes only $0.2\%$ of the total variance at $7\,\mathrm{GHz}$. Accordingly, $u_{\mathrm{tot}}$ is the total standard uncertainty from measurement scatter and known calibration uncertainty within the adopted noise model -- and it is this quantity we report in the main text.

\bibliography{references.bib}
\end{document}